\documentclass[conference]{IEEEtran}
\IEEEoverridecommandlockouts
\usepackage{cite}
\usepackage{listings}
\usepackage{amsmath,amssymb,amsfonts}
\usepackage{algorithmic}
\usepackage{graphicx}
\usepackage{textcomp}
\usepackage{xcolor}
\usepackage{makecell}
\usepackage{tikz}
\usepackage{caption}
\usepackage{float}
\newfloat{lstfloat}{htbp}{lop}
\floatname{lstfloat}{Listing}
\usepackage[T1]{fontenc}
\def\BibTeX{{\rm B\kern-.05em{\sc i\kern-.025em b}\kern-.08em
    T\kern-.1667em\lower.7ex\hbox{E}\kern-.125emX}}

\newcommand\copyrighttext{%
  \footnotesize For the purpose of open access, the author has applied a Creative Commons Attribution (CC BY) license to any Author Accepted Manuscript version arising}
\newcommand\copyrightnotice{%
\begin{tikzpicture}[remember picture,overlay]
\node[anchor=south,yshift=10pt] at (current page.south) {\fbox{\parbox{\dimexpr\textwidth-\fboxsep-\fboxrule\relax}{\copyrighttext}}};
\end{tikzpicture}%
}

\lstdefinelanguage{zig}{
    alsodigit = {.},
    keywords = {fn, pub, extern, const, return, export, var, @intToPtr, @cImport, @cInclude, @import, export, @as}
}    

\begin{document}

\title{Pragma driven shared memory parallelism in Zig by supporting OpenMP loop directives} 

\author{\IEEEauthorblockN{David Kacs}
\IEEEauthorblockA{\textit{EPCC} \\
\textit{University of Edinburgh}\\
Edinburgh, UK \\
d.kacs@sms.ed.ac.uk}
\and
\IEEEauthorblockN{Joseph Lee}
\IEEEauthorblockA{\textit{EPCC} \\
\textit{University of Edinburgh}\\
Edinburgh, UK}
\and
\IEEEauthorblockN{Justs Zarins}
\IEEEauthorblockA{\textit{EPCC} \\
\textit{University of Edinburgh}\\
Edinburgh, UK}
\and
\IEEEauthorblockN{Nick Brown}
\IEEEauthorblockA{\textit{EPCC} \\
\textit{University of Edinburgh}\\
Edinburgh, UK}
}

\maketitle
\copyrightnotice
\begin{abstract}
The Zig programming language, which is designed to provide performance and safety as first class concerns, has become popular in recent years. Given that Zig is built upon LLVM, and-so enjoys many of the benefits provided by the ecosystem, including access to a rich set of backends, Zig has significant potential for high performance workloads. However, it is yet to gain acceptance in HPC and one of the reasons for this is that support for the pragma driven shared memory parallelism is missing.

In this paper we describe enhancing the Zig compiler to add support for OpenMP loop directives. Then exploring performance using NASA’s NAS Parallel Benchmark (NPB) suite. We demonstrate that not only does our integration of OpenMP with Zig scale comparatively to Fortran and C reference implementations of NPB, but furthermore Zig provides up to a 1.25 times performance increase compared to Fortran. 
\end{abstract}


\section{Introduction}
As the High Performance Computing (HPC) community moves further into the exascale era, a key question is around the languages that are being used to program our ever increasingly complicated supercomputers. Whilst there has been some progress made in this area, traditional languages such as Fortran and C still make up the majority of code that is run on these machines. 

One alternative is Zig, which is a systems programming language created by Andrew Kelly in 2016. Designed to be fast and safe, a vibrant community has grown up around this language in recent years. Whilst there has been some tangential adoption of Zig in the HPC space, for instance Cerebras's CSL programming technology used to write codes for their Wafer Scale Engine (WSE) is built upon Zig, Zig is yet to gain traction more widely in HPC. One of the reasons for this lack of adoption is that support in the language for common HPC programming technologies is missing. Given the extensive interoperability features in Zig for C, leveraging MPI is in fact relatively straightforward, however support for pragma driven shared memory parallelism which is ubiquitous in HPC codes requires modifications to the compiler and is missing.

Zig leverages the LLVM ecosystem as part of its compiler and runtime, and in this paper we explore an enhancement to the Zig compiler to support pragma driven shared memory parallelism via OpenMP loop directives. Ultimately calling into LLVM's OpenMP runtime library, we describe the modifications required to support OpenMP loop directives before comparing performance for kernels in NASA's NAS Parallel Benchmark suite, NPB, between C, Fortran and Zig. This paper is structured as follows; after describing the background to this work in Section \ref{sec:bg}, we then explore our enhancements to Zig in Section \ref{sec:openmp-zig} in order to support OpenMP loop directives. Section \ref{sec:experiment-approach} highlights the approach we have taken to evaluate this work, including an exploration of integrating Zig, C and Fortran, before investigating the performance of Zig and OpenMP for these benchmarks in Section \ref{sec:results}. Lastly, Section \ref{sec:conclusions} draws conclusions and discusses further work. 

The main contributions of this paper are:
\begin{itemize}
  \item A description of how OpenMP loop directives and the Zig compiler can be integrated together.
  \item Exploration of how to integrate Zig with Fortran codes, which has never been done before, potentially enabling Zig to be leveraged as part of a much larger traditional code base.
  \item A performance comparison of Zig against Fortran and C across three HPC benchmarks, both in terms of the speedup obtained when threading but also the runtime itself. We demonstrate that Zig performs well against these other languages and is a viable option for HPC.
\end{itemize}

\section{Background and related work}
\label{sec:bg}
LLVM is a collection of tools and libraries that in part is used to generate and manipulate a common internal representation known as  LLVM-IR\cite{lattner_2012}. There are a rich set of backends built upon LLVM that generate machine code from this internal representation and target a variety of hardware including CPUs, GPUs and FPGAs. In addition to providing frontends such as \emph{Clang} and \emph{Flang}, for C and Fortran, as part of the main repository, LLVM is also used by many other popular programming languages, including Swift\cite{swift_2023}, Rust\cite{rustc_2023} and Zig \cite{kelly-zig}.

LLVM also provides its own OpenMP library, where OpenMP provides pragma driven shared memory parallelism via multithreading and is one of the most popular programming technologies used in HPC. The core functionality of OpenMP is implemented through directives and consequently modifications to the compiler are required to support these pragmas. The OpenMP standard \cite{omp-api} specifies how programmers in C, C++ and Fortran should interact with the technology, where compiler directives are specified as pragmas in C and C++ and as special comments in Fortran. This sequence of tokens which defines the start of the directive is referred to as a sentinel.

\subsection{Zig}
Zig is a systems programming language created by Andrew Kelly in 2016 with the goal of being an optimal, safe and readable alternative to C \cite{kelly-zig}. The design of Zig is focused on reducing program execution time, whilst striving for a higher level of safety and an improved programming experience compared to C. Zig leverages the LLVM compiler infrastructure \cite{llvm} for its code generation, which enables it to take advantage of its optimisation features \cite{llvm-passes}. The use of LLVM also enables Zig to support a large number of CPU architectures and operating systems, the the objective being to support all targets that are supported by LLVM \cite{zig_llvm_bindigs_2023}. 

Zig provides a number of safety features to improve the software development experience. The key ones being a stronger type system and improved  static analysis compared to C, and optional runtime safety checks enabled when  using the compiler's debug mode. Static analysis features can aid the programmer in preventing common bugs such as dereferencing a null pointer or truncation and rounding errors associated  with integer and floating point casting. For example, in C for \emph{int *ptr = 0}, dereferencing and reading from \emph{ptr} is legal but would likely cause a segmentation fault at runtime. The two code examples in Listing \ref{lst:zig-safety} illustrate how this would be represented in Zig. However, neither of these examples compile, the first attempts to assign an integer literal to a pointer and such implicit conversions are prohibited by the Zig type system. The second example in Listing \ref{lst:zig-safety} uses the built in \emph{@intToPtr} Zig function to perform an explicit integer to pointer cast, however this fails because in Zig only nullable pointers may be assigned the value zero.

\begin{lstlisting}[frame=lines, label=lst:zig-safety, numbers=left, caption=Examples how in Zig one represent dereferencing and reading from \emph{prt} based upon \emph{int *ptr = 0} in C however neither of these examples will compile in Zig due to safety provided by the language]
var ptr: *i32 = 0;
_ = ptr.*;
// ---------------------------------------
var ptr: *i32 = @intToPtr(*i32, 0);
_ = ptr.*;
\end{lstlisting}

Errors which cannot be identified at compile time may be flagged at runtime through safety checked undefined behaviour. Zig provides two execution modes for compiled code, a production and debug mode. In debug mode additional code is inserted into the executable, for instance checking whether indexing outside the bounds of an array has occurred or integer overflow. A runtime error is then raised if such situations occur. By contrast, for performance reasons, production mode does not provide such safety and such undefined behaviour will go unchecked. The approach being that programmers develop their code using debugging mode, and then switch to production mode once they are confident that it is mature.

One of the design goals of Zig was to provide interoperability with existing C code bases \cite{kelly-zig-c-abi}. This enables one to leverage C libraries and framework, such as MPI \cite{mpi}, as well as being able to gradually replace C with Zig in a project. To achieve this interoperability, Zig provides a method for both calling C functions and allowing Zig functions to be called from C. Figure
\ref{lst:call-c} shows the C standard library function \emph{puts} being invoked from Zig. 

\begin{lstlisting}[frame=lines, label=lst:call-c, numbers=left, caption=An example of calling a C function from Zig]
extern fn puts(s: [*:0]const u8) c_int;

pub fn main() void {
    // calling puts
    _ = puts("hello world");
}
\end{lstlisting}

Listing \ref{lst:call-zig} illustrates a function, \emph{add}, being exported for use in C. This function can be accessed in C by either compiling to an object file or a static library, and linking against a C program. Furthermore, a C header file can be automatically generated by Zig which contains the signatures of all exported functions.

\begin{lstlisting}[frame=lines, label=lst:call-zig, numbers=left, caption=An example of exporting a Zig function to be used in C]
pub export fn add(a: c_int, b: c_int) c_int{
    return a + b;
}
\end{lstlisting}

The Zig compiler also provides tools for converting C source code to Zig which can be used to accelerate the process of converting a project, or parts of it, to Zig. This mechanism is also utilised by the compiler to automatically parse C header files and import their functions, structures and constants. Figure \ref{lst:cimport} shows a modified example from Figure \ref{lst:call-c} with  the explicit declaration of the function \emph{puts} being replaced by an automatic  translation of the \emph{stdio.h} header of the C standard library.

\begin{lstlisting}[frame=lines, label=lst:cimport, numbers=left, caption=An example of importing a C header and invoking a function from it by Zig]
// Importing the C stdio.h header
const stdio = @cImport(@cInclude("stdio.h"));

pub fn main() void {
    // calling the puts function from the stdio.h header
    _ = stdio.puts("hello world");
}
\end{lstlisting}

\section{Supporting OpenMP in Zig}
\label{sec:openmp-zig}
LLVM provides its own OpenMP runtime library, and our goal in this work has been to call into functions provided by that library to provide pragma driven shared memory programming in Zig. In this section we explore the methodology adopted to connect a programmer's Zig source code, annotated with OpenMP pragmas, to LLVM's runtime library. All modifications were applied to the Zig compiler version 0.10.1.

\subsection{Tokenisation and parsing}

\begin{figure}[htb]
\centering
 \includegraphics[width=\columnwidth]{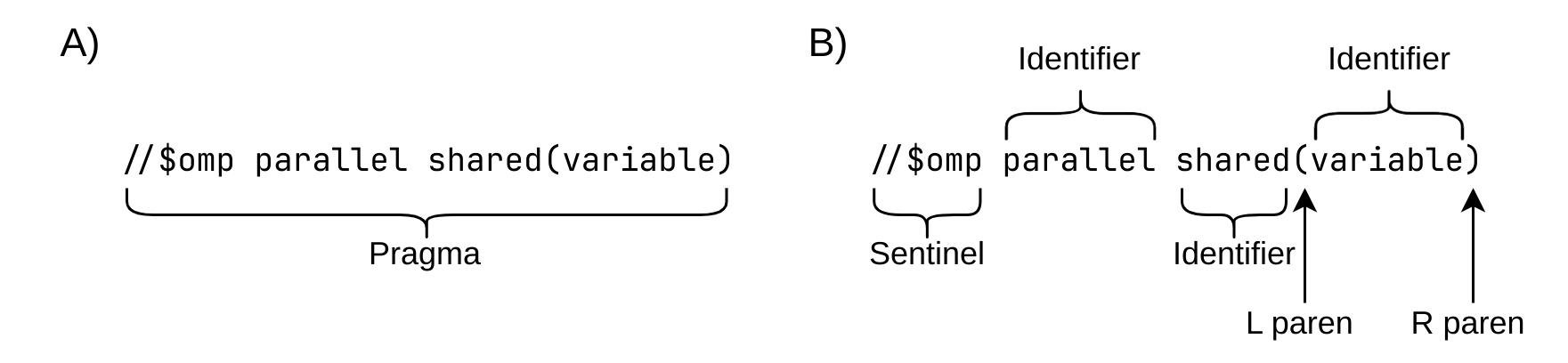}
\caption{Illustration of a choice of parsing, either A) parsing the entire pragma as a single token or B) separating the pragma into multiple tokens}	
\label{fig:pragma}
\end{figure}

As discussed in Section \ref{sec:bg}, OpenMP relies on pragmas to specify how a program should be parallelised, however pragmas are not an existing feature of Zig. Consequently they had to be added as a new kind of statement and the decision we made was to implement pragmas as special comments, which is similar to how they are supported in Fortran. The first step in the Zig compilation pipeline is the tokeniser, and the main choice that had to be made was whether to parse the whole pragma as a single token, or parse each component as separate tokens, options A) and B) in Figure \ref{fig:pragma} respectively. The decision was made to leverage the existing tokeniser infrastructure by tokenising the sentinel and then tokenising the rest of the pragma as regular code as if the sentinel was not present. This is possible as the pragma consists entirely of tokens used by Zig itself.

The Zig tokeniser provides a mechanism for tokenising keywords. Consequently, the initial plan was to utilise this mechanism to parse OpenMP directives and clauses, such as \emph{parallel} or \emph{default} as keywords. However, this was not possible, because in Zig keywords may not be used as identifiers, and adding these would break compatibility with existing codes. The solution was therefore to store OpenMP keywords as identifiers and differentiate these from regular identifiers during parsing.

After tokenisation the next step is parsing and this generates an Abstract Syntax Tree (AST) from the tokens. Pragmas should be treated the same way as any other statement, and the core of the Zig parser revolves around the \emph{eatToken} function. This function accepts an enumeration representing the type of a token, referred to as a token tag. If the next token matches the tag, is returned and the parser advances to the next token, otherwise \emph{null} is returned. This function, however, could not be used in its original form, due to OpenMP keywords not being assigned unique tags. Consequently, a new set of tags was added to represent the different OpenMP keywords, with a hash map of strings to keyword tokens used to identify whether a string is a keyword. We modified the \emph{eatToken} function to accept both existing and new tags, and parse the identifier tag accordingly if an OpenMP keyword tag was used. 


Each OpenMP directive is provided with an AST node tag, with the clauses stored as node data. Clause data is stored in the \emph{extra\_data} array which is an array of 32 bit integers and already part of the Zig compiler used to annotate miscellaneous data about nodes in the AST. Whilst the Zig compiler provides a mechanism for storing structures in this array, every element of the structure must be a 32 bit integer. Consequently, all clause data has to be representable in this form where all clauses are stored in a single data structure, with the integer representation within this data structure varing between clauses.

\subsubsection{Handling list clauses}
\begin{figure}[htb]
\centering
 \includegraphics[width=\columnwidth]{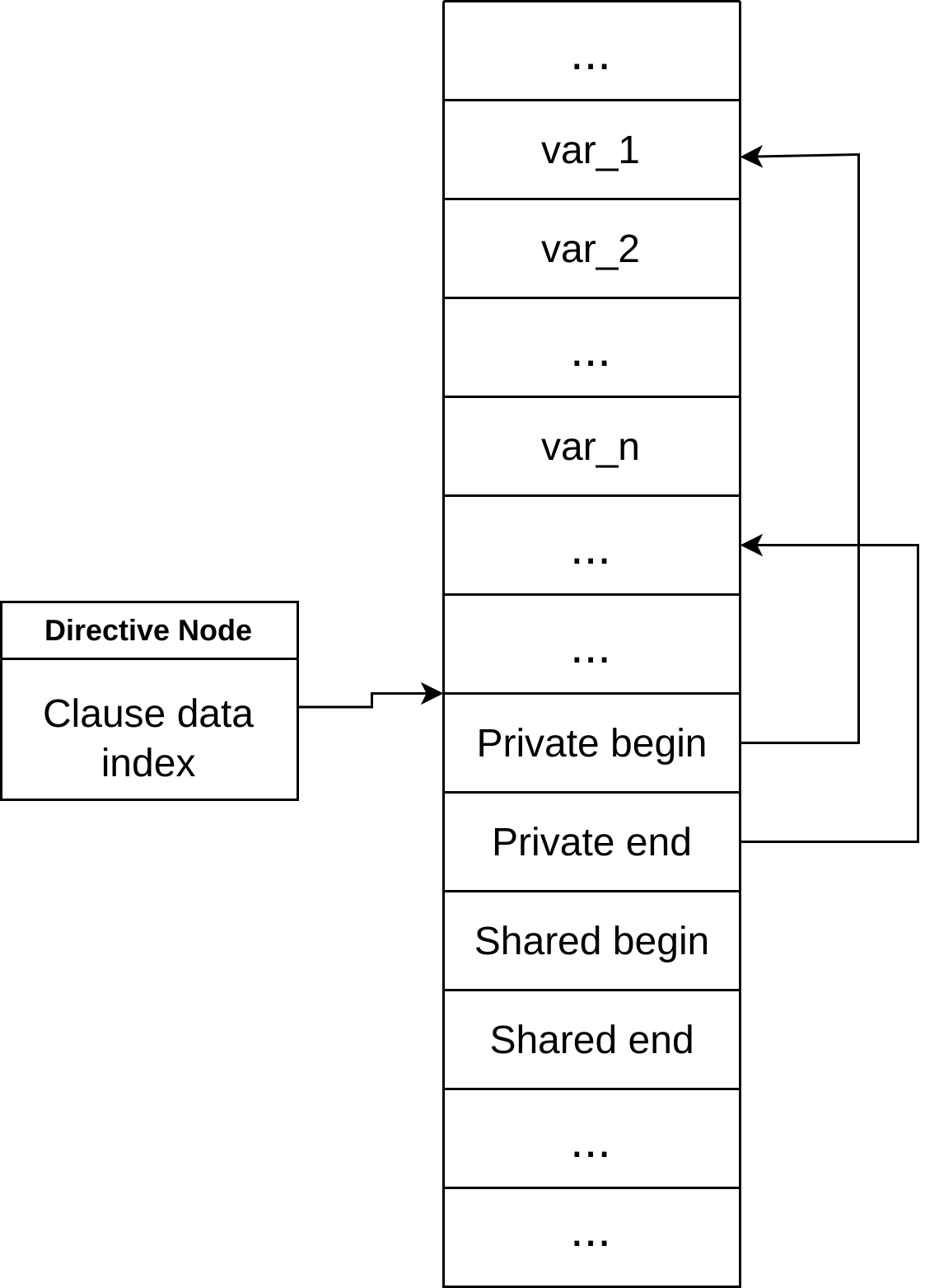}
\caption{Private variables stored in the \emph{extra\_data} array}	
\label{fig:private}
\end{figure}

The \emph{private}, \emph{firstprivate} and \emph{shared} clauses are defined as a list of identifiers. After obtaining the AST node indices of each identifier, these are stored contiguously in the \emph{extra\_data} array, with the beginning and end indices of these slices being stored in clauses. Figure \ref{fig:private} provides an example of this for the \emph{private} clause, and illustrates how a directive node contains an index into the \emph{extra\_data} array denoting the start of the clauses structure. 

\subsubsection{Handling of packed clauses}
The storage size for the non-list clauses is know statically and-so these can be stored in a single structure. By making this structure packed, it is possible for this to be interpreted as a 32 bit integer and stored in the \emph{extra\_data} array. Using this approach means that extracting all data can be extracted by reading a single index of the array without further indirection. For instance, the loop schedule is stored as a 3 bit enumeration representing the schedule type followed by a 29 bit integer representing the chunk size, which allows for a maximum chunk of 536870912 iterations. Because the chunk size must be greater than 0 \cite{omp-api}, the value 0 is used represent no chunk size having been specified.

Several clauses can be represented by fewer than 32 bits, and these are grouped into a single packed structure. For example, the \emph{default} clause is represented by a 2 bit enumeration and the \emph{nowait} clause represented by a boolean which is a single bit in the packed structure. The \emph{collapse} clause is represented by 4 bits, as it is unlikely that a user would wish to collapse more than 16 loops. 

\subsection{Code generation}

Once OpenMP pragmas have been tokenised and parsed, the next step is to then use these in the generation of code. A typical compiler that supports OpenMP will add calls to the OpenMP runtime in the place of directives. Such a replacement requires augmenting the program, which is represented as an AST, during compilation.

Our initial attempt was to directly modify the AST and inject the required OpenMP calls. However, in Zig there is a strict connection between AST nodes and the original source code, meaning that the addition of arbitrary additional nodes is not feasible. Based upon this, a variation was also attempted, which involved pre-pending a function and structure definition to the start of the programmer's source code before it was parsed in order for these to be used as a template that could be copied during code generation, for all OpenMP function and structure instantiations. We felt that this would be a sensible workaround since two AST nodes are allowed to refer to the same location in source code. However, finding the locations of these templates and propagating them through the AST during compilation was found to be unfeasible based upon the current compiler design of Zig. 

Consequently, we adopted a preprocessor based approach, which has the advantage that the new code can be easily synthesised without the need for manually ensuring that each token and AST node references a fixed location in the source file. There are a couple of challenges associated with adopting a preprocessing approach however, mainly because Zig was not designed to incorporate such a step. Firstly, all unused function arguments and variables that are not in the global scope must be explicitly discarded which means that all variables should be generated only if it is known that they will be used. The second challenge is the lack of semantic context, such as variable types and their usage during the preprocessing step and this is discussed in more detail in Section \ref{sec:rewriting}.

The decision was made to add the preprocessor as an integral part of the Zig compiler which has several benefits. Firstly, this enables the preprocessor to reuse the parsing infrastructure built into the Zig compiler, and secondly by executing the preprocessor immediately after a file is loaded then it is possible for the compiler's caching mechanism to continue working with no modifications required.

Our preprocessor runs in several passes, each focusing on replacing separate OpenMP constructs. The pseudo code of our overarching algorithm is sketched in Listing \ref{lst:algorithm}, where it can be seen that, for example, all parallel regions are replaced before worksharing loops. Consequently, nested constructs do not require special handling in the preprocessor as long as they are of different types. The construct \emph{<<adjust source offset>>} can be seen in Listing \ref{lst:algorithm}, this is because nodes are expressed as offsets into the source code listing and consequently the offset at which the code is modified has to be adjusted after each replacement. The pseudo code also shows a payload being created, via \emph{create-payload}, for each node that is replaced. This payload contains the information required to perform such a replacement, for example each directive requires the location in the source code where the replacement is to be made, as well as information specific to the directive. 

\begin{lstlisting}[language=Matlab, frame=lines, label=lst:algorithm, numbers=left, caption=Psuedo code of our preprocessor algorithm to replace OpenMP pragmas and clauses]
FUNCTION preprocess (source, step)

  ast          := parse-source-into-ast(source)
  replacements := empty-list

  FOREACH node IN ast DO
    IF node IS OpenMP-node AND
       <<node matches current step>> THEN
      append(replacements, create-payload(node))
    END
  END

  IF step = parallel THEN
    FOREACH replacement IN replacements
      <<perform parallel region replacement>>
      <<adjust source offset>>
    END
  ELSE IF step == while THEN
    FOREACH replacement IN replacements
      <<perform worksharing loop replacement>>
      <<adjust source offset>>
    END
  END

  IF <<is last step>> THEN
    RETURN source
  ELSE
    RETURN preprocess(source, step)

END
\end{lstlisting}

\subsubsection{Handling parallel regions}
\label{sec:reduction}
Most compilers express OpenMP parallel regions using function outlining, where a function is generated whose body contains the contents of the parallel region \cite{gomp-parallel}. Variables which are accessed by this parallel region, such as those shared by default or explicitly captured using the \emph{shared}, \emph{firstprivate} or \emph{reduction} clauses, are then passed to the function as arguments. A pointer to this outlined function is then passed to the OpenMP library's runtime function which will invoke it on each thread. For example. LLVM's OpenMP API performs this via the \emph{\_\_kmpc\_fork\_call} function.

We adopted this approach and variables passed to the outlined function are provided as arguments to the OpenMP runtime library function \emph{\_\_kmpc\_fork\_call}. This then forwards them as part of the callback to our outlined function. The \emph{\_\_kmpc\_fork\_call} function is variadic, meaning that it accepts a variable number of arguments. This provided us with flexibility around how arguments are passed to it. Our design was to pass arguments in one of three groups, each represented as \emph{?*anyopaque} pointers which as Zig's equivalent to C's void *. These three arguments groups point to structures containing variables provided as part of the \emph{firstprivate}, \emph{shared} and \emph{reduction} clauses respectively. 

In the outlined function once \emph{?*anyopaque} pointers are cast back to their original types, a variable is created for each member of these structures and initialised with the value from the structure. For the \emph{firstprivate} clause this value is simply the value of the variable from the scope surrounding the parallel region. For \emph{shared} this is a pointer to a variable, but also requires each shared variable access to be subsequently rewritten as pointer accesses. Private variables are simple, as they are simply defined in the outlined function.

Reduction operations are more complex, and were implemented by creating a value using Zig's standard \emph{atomic} type. A reduction structure is created with pointers to these atomic values and this is passed to the outlined function callback in the same manner as other variables. The outlined function then creates a separate variable for each reduction variable, which are then initialised using the initial value held in the reduction variable and this initialisation is required to conform with the OpenMP standard \cite{omp-api}. Atomic read-modify-write operations are defined on the atomic type to ensure modification of the atomics in a thread safe manner.

However, this approach is limited by the atomic operations that are supported by Zig. At present, only addition, subtraction, minimum, maximum, the binary AND, OR, NAND and XOR, and compare-and-swap are provided. Notably, multiplication and logical AND and OR not supported. We implement these missing reduction operations using the compare-and-swap (CAS) loop algorithm \cite{sutter-cas}, and a pseudo code sketch of our algorithm for implementing multiplications is provided in Listing \ref{lst:cas}. The function \emph{compare-and-swap} compares the values of \emph{atom} to that of \emph{old}. If these values are equal then the value held in \emph{atom} is set to that of \emph{new}. Because this operation is atomic the value of \emph{atom} cannot change throughout the exchange. The \emph{compare-and-swap} function returns two values, \emph{exchange-success} and \emph{actual-value}. The former indicates whether the value of \emph{atom} has been updated, and the later the resulting value, with the loop continuing until \emph{atom} is updated.

\begin{lstlisting}[language=Matlab, frame=lines, label=lst:cas, numbers=left, caption=Psuedo code of our algorithm using Compare And Swap (CAS) to implement multiplication reduction]
atom    := <<value to be updated>>
operand := <<value to update it with>>

old := atomic-load(atom)
new := old * operand

WHILE TRUE DO
  exchange-success, actual-value :=
    compare-and-swap(&atom, old, new)
  IF exchange-success THEN
    BREAK
  ELSE
    old = actual-value
    new = old * operand
  END
END
\end{lstlisting}

\subsubsection{Handling worksharing loops}

Unlike parallel regions, worksharing loops do not require an outlined function. The \emph{Clang} OpenMP API provides two distinct strategies for implementing worksharing loops, the first one used by the static schedule, which is the \emph{\_\_kmpc\_for\_static\_*} function and the second for dynamic, guided and runtime schedules which is the \emph{\_\_kmpc\_dispatch\_*} function. Both of these require the loop's upper bound, lower bound, increment and comparison operator have to be determined. We take the comparison operator directly from the condition of the Zig while loop, and the lower bound is determined by the initial value of the loop counter variable. The upper bound is taken from the right-hand-side of the comparison operator, and the increment is determined from the value on the right hand side of the increment operator in the continuation expression. 

These two functions are stared here, because variants of both \emph{\_\_kmpc\_for\_static\_*} and  \emph{\_\_kmpc\_dispatch\_*} are provided. For example, with the static clause \emph{\_\_kmpc\_for\_static\_init} will undertake the loop iterations with \emph{\_\_kmpc\_for\_static\_fini} called by every thread to finalise the loop upon completion. In addition to these functions, to handle dynamic loops a \emph{\_\_kmpc\_dispatch\_next} function is also provided to process the next batch of iterations if such are available. Furthermore, the \emph{\_\_kmpc\_dispatch\_init} function also accepts the schedule type, which may be one of \emph{kmp\_sch\_dynamic\_chunked}, \emph{kmp\_sch\_guided\_chunked} \emph{kmp\_sch\_runtime} which corresponds to \emph{dynamic}, \emph{guided} and \emph{runtime} schedules respectively.

\subsubsection{Variable rewriting}
\label{sec:rewriting}
Our preprocessor aims to leverage existing variable names and expressions where possible, for example reusing the same variable names when unpacking \emph{private} and \emph{firstprivate} variables in the outlined function. However, this is not always possible, for example \emph{shared} variables must be rewritten to be accessed via pointers or worksharing loop reduction temporaries which may not share their names with the shared variable that they are being reduced into.

Such substitutions are made more difficult due to the lack of semantic context at the time of preprocessing. However, Zig's relatively simple grammar \cite{zig-grammar} and lack of shadowing makes it possible to perform these rewrites only using the AST. The use of
variables can be determined by comparing the values of their identifiers, where two identifiers in the same scope will always refer to the same entity as long as neither is preceded by a period.

\subsection{Wrapping OpenMP runtime functionality}

The OpenMP standard describes a set of runtime functions that must  be provided by a conforming OpenMP implementation as part of the runtime library. These functions are intended to be called directly by the user and can be identified by their \emph{omp\_} prefix, such as \emph{omp\_get\_thread\_num} or \emph{omp\_get\_num\_threads}. To make these functions available to Zig programmers we added an omp namespace and included it in the standard library. All function declarations were translated from C to Zig using the Zig compiler's \emph{translate-c} feature, and these translated function declarations were then re-exported with the \emph{omp\_} prefix removed which was redundant once the functions were placed in the namespace. Listing \ref{lst:thread_id} illustrates using this approach in Zig to obtain the thread's id.

\begin{lstlisting}[language=zig, frame=lines, label=lst:thread_id, numbers=left, caption=Illustration of obtaining the thread ID in Zig using our OpenMP library wrapper]
const omp = @import("std").omp;
const thread_id = omp.get_thread_num();
\end{lstlisting}

In addition to the standard OpenMP API that is made available to the Zig programmer, an internal OpenMP API was required to be used by our preprocessor to drive the mapping of OpenMP pragmas to the LLVM OpenMP runtime library. These functions and constants are not specified by the OpenMP standard, but instead the the runtime, \emph{libomp}, in this case  the LLVM OpenMP library implementation \cite{llvm-libomp}. These function declarations were translated the same way as we did for the standard OpenMP functions, but placed within a \emph{.omp.internal} namespace. Unlike the standard API, the prefix was not removed when exporting these functions, as they are not intended to be used by programmers directly. In addition to exposing the functions provided by the LLVM OpenMP runtime, the .omp.internal namespace also contains various helper utilities that we developed and are used by our preprocessor. These include generic wrappers for the \emph{\_\_kmpc\_dispatch\_*} and \emph{\_\_kmpc\_for\_static\_*} families of functions as well as the CAS loop reduction described in Section \ref{sec:reduction}.

\section{Experimental approach}
\label{sec:experiment-approach}
All benchmarks conducted in this paper are run over a single node of the Cray-EX ARCHER2 supercomputer. Each node consists of two 64-core AMD EPYC 7742 processors with 32KB of L1 data cache, 32KB of L1 instruction cache and 512KB of L2 cache per core as well as 16.4MB of L3 cache shared between groups of four cores. The OpenMP runtime used for all benchmarks was \emph{libomp} based on LLVM version 13.0.0 which is used in the the AMD Optimising C and Fortran compilers (AOCC). Reference implementations of C and Fortran benchmarks were compiled with AOCC compilers utilising the same OpenMP runtime, \emph{Clang} and \emph{Flang} respectively. The Fortran code that integrates with the Zig versions of our benchmarks was compiled with the GNU \emph{gfortran} compiler version 7.5.0, and all C code was compiled with AMD \emph{Clang} version 13.0.0. The versions of AOCC and \emph{gfortran} used in the benchmarks were the latest available on the ARCHER2 platform at the time benchmarking was conducted. Each benchmark was ran 5 times for each thread count, and the mean of these 5 runs is reported here. The execution time was measured using the internal timers provided within the reference implementations. 

As no established Zig HPC benchmarks presently exist the decision was made to use benchmarks created for established HPC languages, such as Fortran and C, and convert these to Zig. To convert from C to Zig, the \emph{translate-c} subcommand has been provided as part of Zig. However, this utility has several drawbacks, firstly it ignores all C pragmas meaning all OpenMP specific information is lost. The second limitation is that \emph{translate-c} executes the C preprocessor prior to translating the code to Zig, meaning that all headers included with \emph{\#include} are also translated and appear in the resulting Zig. An example C code is illustrated in Listing \ref{lst:c-benchmark} which defines a function that returns a preprocessor-defined constant. It can be seen in the translated function of \ref{lst:zig-benchmark} that whilst \emph{CONSTANT} has been defined, this is not used and instead the expanded value is returned. Consequently, whilst this tool was helpful these limitations meant that we still had to manually port parts of the code ourselves and undertake extensive verification.

\begin{lstlisting}[language=C, frame=lines, label=lst:c-benchmark, numbers=left, caption=C program that will be translated with the \emph{translate-c} subcommand]
#define CONSTANT (37 + 5)

int foo(void) { return CONSTANT; }
\end{lstlisting}

\begin{lstlisting}[frame=lines, label=lst:zig-benchmark, numbers=left, caption=Zig program translated from the C code using the \emph{translate-c} subcommand]
pub export fn foo() c_int {
    return @as(c_int, 37) + @as(c_int, 5);
}
pub const CONSTANT = @as(c_int, 37) + @as(c_int, 5);
\end{lstlisting}

The Zig compiler does not provide a Fortran equivalent to \emph{translate-c}, resulting in all Fortran code being ported manually. However, there are several major differences between Zig and Fortran, with the most significant being 1-indexed arrays and inclusive DO loop upper bounds in Fortran but not in Zig. Consequently, all array indices and loop lower bounds have to be adjusted in such ports which adds complexity.

Whilst it has never been done before, the process of invoking Fortran procedures from Zig is similar to calling C functions. Procedures are declared in Zig as functions with C linkage, with all argument types changed to pointers. Furthermore, to conform with LLVM's name mangling scheme an underscore has to be appended to the end of the function name. It is also possible to call Zig functions from Fortran, but one must again be careful of the name mangling scheme. An example of this is that only GNU \emph{gfortran} provides predictable names for global variable entries and it is this reason that, in our ports of these benchmarks described in Section \ref{sec:results}, when integrating with Fortran we compile the remaining Fortran code using GNU \emph{gfortran}. A more robust approach would be to engage with the standards body and develop an enhancement similar to Fortran 2003's C interoperability features \cite{f2003-std}.

\section{Results and evaluation}
\label{sec:results}
In this work we leverage kernels from the NAS Parallel Benchmarks (NPB) suite \cite{npb}, which comprises various kernels based on algorithmic patterns commonly found in Computational Fluid Dynamics (CFD) applications. 

\subsection{Conjugate Gradient (CG)}

Conjugate Gradient (CG) was the first benchmark we selected and this utilises a large number of OpenMP features supported by our approach. We ported the \emph{conj\_grad} subroutine, which accounts for around 95\% of the runtime, from Fortran into Zig. This subroutine includes parallel and worksharing directives, \emph{private}, \emph{shared} and \emph{firstprivate} variable sharing clauses, the \emph{nowait} clause, as well as reductions on both the parallel region and the worksharing loops. Furthermore, this kernel is also representative of iterative algorithms which form a large class of workloads in HPC.

\begin{figure}[htb]
\centering
 \includegraphics[width=\columnwidth]{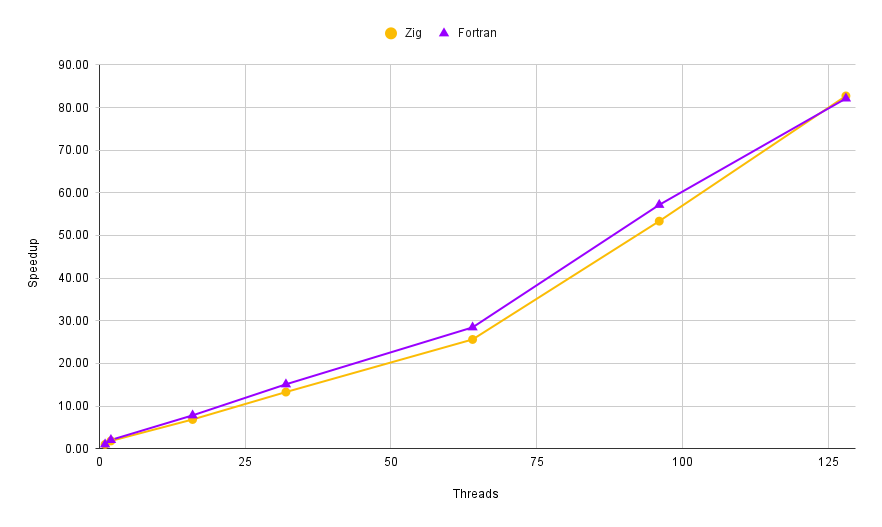}
\caption{Speedup against number of threads for the CG benchmark (class C) for both our approach in Zig and the Fortran reference implementation}	
\label{fig:cg}
\end{figure}

Figure \ref{fig:cg} reports the speedup for the CG kernel, running at problem size class C, when strong scaling across varying numbers of threads. Both languages generally follow Amdahl's Law up to 64 threads, but perform significantly better than expected when run over 96 and 128 threads. This appears to be an inherent property of the algorithm, as both our Zig port and the reference implementation follows an almost identical speedup curve, demonstrating very similar performance between OpenMP in Fortran and OpenMP in Zig for this benchmark.

\begin{table}[htb]
    \centering
    \caption{Runtime of Zig and Fortran NPB CG benchmark (class C) for different number of threads when strong scaling}
    \label{tab:cg-runtime}
    \begin{tabular}{|c|c|c|}
    \hline           
     \textbf{Number of threads} & \textbf{Zig runtime (s)} & \textbf{Fortran runtime (s)} \\
      \hline
    1 & 149.40 & 170.17 \\
    2 & 82.34 &  83.35 \\
    16 & 21.85 &  21.80 \\
    32 & 11.26 &  11.28 \\
    64 & 5.83 &  5.98 \\
    96 & 2.80 & 2.98 \\
    128 & 1.81 &  2.07 \\
    \hline
    \end{tabular}
\end{table}

Table \ref{tab:cg-runtime} reports the runtime of the CG benchmark in Zig and Fortran, and it can be seen that the Zig version is 1.15 times faster than the Fortran code on a single core and then performance remains roughly equal for all other thread counts, although Zig tends to be marginally faster than Fortran.

\subsection{Embarrassingly Parallel (EP)}

The Embarrassingly Parallel (EP) kernel focuses on compute performance alone, with no synchronisation required between the threads and an efficient memory access pattern. We ported the entirety of the code, apart from the timing and verification routines, from Fortran into Zig. This kernel utilises \emph{private} and \emph{firstprivate} variable sharing clauses, as well as a parallel region reduction. Additionally, \emph{threadprivate} and \emph{atomic} directives are also used.

\begin{figure}[htb]
\centering
 \includegraphics[width=\columnwidth]{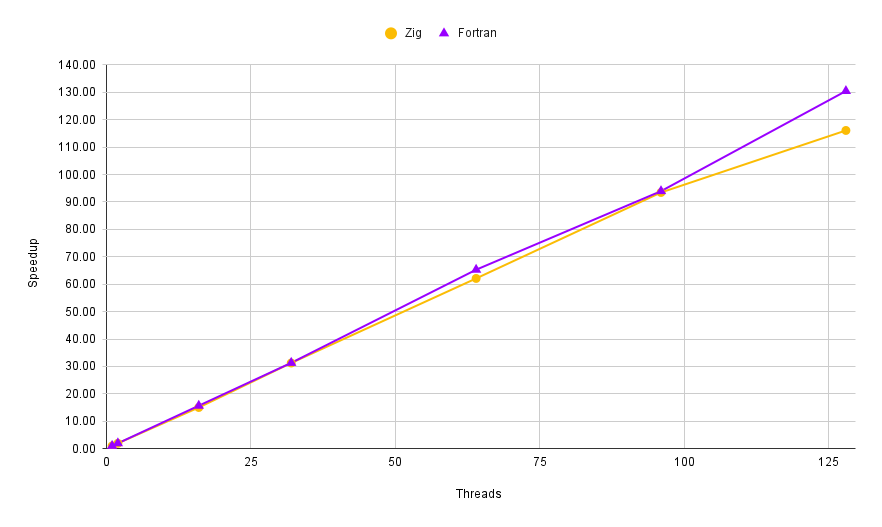}
\caption{Speedup against number of threads for the EP benchmark (class C) for both our approach in Zig and the Fortran reference implementation}	
\label{fig:ep}
\end{figure}

Figure \ref{fig:ep} illustrates speed up for the EP benchmark for both the Zig port and Fortran reference implementation versions when strong scaling at problem size class C. It can be seen that the speedup is directly proportional to the thread count for both the Zig port and the reference implementation, and this is unsurprising given that the algorithm requires no communication between threads. The outlier is at 128 threads, where the Fortran reference implementation delivers a  speedup in excess of 128 times, meaning that the benchmark is benefiting from super linear scaling and  this is not observed in the Zig port. This is likely due to better cache utilisation for the Fortran version across a larger number of threads as the problem size is reduced per thread. 

\begin{table}[htb]
    \centering
    \caption{Runtime of Zig and Fortran NPB EP benchmark (class C) for different number of threads when strong scaling}
    \label{tab:ep-runtime}
    \begin{tabular}{|c|c|c|}
    \hline           
     \textbf{Number of threads} & \textbf{Zig runtime (s)} & \textbf{Fortran runtime (s)} \\
      \hline
    1 & 147.66 & 185.26 \\
    2 & 76.17 & 94.90 \\
    16 & 9.84 &  11.83 \\
    32 & 4.72 & 5.92 \\
    64 & 2.29 & 2.84 \\
    96 & 1.57 & 1.97 \\
    128 & 1.36 &  1.42 \\
    \hline
    \end{tabular}
\end{table}

Table \ref{tab:ep-runtime} reports the runtime of the Zig port and Fortran reference implementation of the EP benchmark, when strong scaling, and it can be seen that the Zig version is on average 1.2 times faster than the reference implementation. This is similar to the CG benchmark and surprised us given Fortran's popularity for these scientific workloads. Even though the Fortran version scales better at 128 cores, it is still executing more slowly than the Zig version of the benchmark.

\subsection{Integer Sort (IS)}
The Integer Sort (IS) kernel comprises indirect memory accesses and is designed to pressurise the memory subsystem. This kernel leverages the \emph{private} and \emph{firstprivate} sharing directives, as well as using a \emph{static,1} schedule. The major difference between the IS benchmark and others considered in this paper is that it is written in C, and we ported the \emph{rank} function which accounts for around 70\% of the total runtime into Zig.

\begin{figure}[htb]
\centering
 \includegraphics[width=\columnwidth]{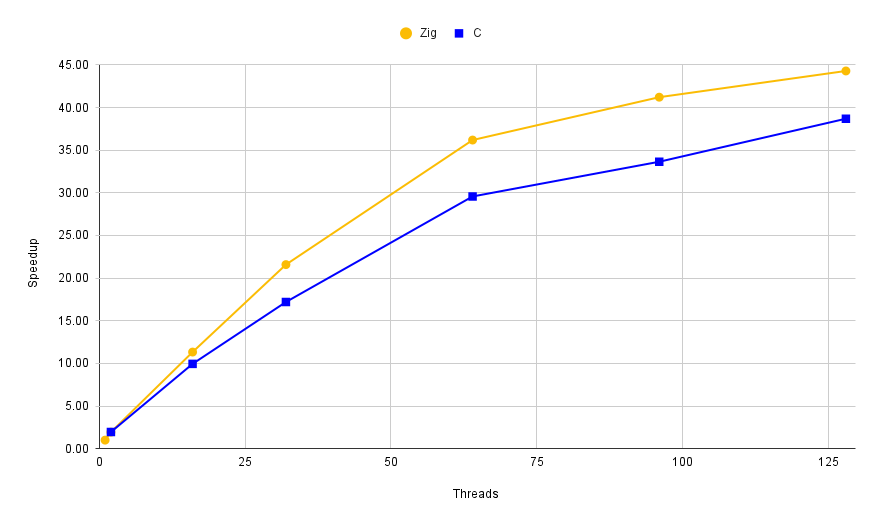}
\caption{Speedup against number of threads for the IS benchmark (class C) for both our approach in Zig and the C reference implementation}	
\label{fig:is}
\end{figure}

Figure \ref{fig:is} reports the speed up at varying numbers of threads at the class C problem size for the Zig port compared to the C reference implementation when strong scaling. It can be seen that both these versions of the benchmark follow a very similar scaling pattern across the entire thread count, however the Zig version initially scales better and-so at larger numbers of threads is offset to provide a greater speed up.

\begin{table}[htb]
    \centering
    \caption{Runtime of Zig and Fortran NPB IS benchmark (class C) for different number of threads when strong scaling}
    \label{tab:is-runtime}
    \begin{tabular}{|c|c|c|}
    \hline           
     \textbf{Number of threads} & \textbf{Zig runtime (s)} & \textbf{Fortran runtime (s)} \\
      \hline
    1 & 11.87 & 9.29 \\
    2 & 6.12 & 4.76 \\
    16 & 1.05 & 0.93 \\
    32 & 0.55 & 0.54 \\
    64 & 0.33 & 0.31 \\
    96 & 0.29 & 0.28 \\
    64 & 0.27 &  0.24 \\
    \hline
    \end{tabular}
\end{table}

Table \ref{tab:is-runtime} reports the runtime for the IS benchmark when strong scaling, and it can be seen that in contrast to the Fortran benchmarks, for this C implemented benchmark, it is the C version that performs best over one thread. Whilst this difference is significant when running in serial, better scaling of the Zig implementation closes the gap and at a greater number of threads performance is very similar between the two languages.

\section{Conclusions and further work}
\label{sec:conclusions}
In this paper we have explored enhancing Zig with loop sharing constructs of OpenMP. Designed initially as a systems programming language, and leveraging the LLVM ecosystem, a major feature of the language is to provide performance and safety, which makes it a very interesting potential future programming language for HPC. Whilst the ability to call C functions in Zig means that integration with MPI is fairly trivial, supporting the pragma based approach of OpenMP requires additional work on the compiler, yet it is crucially important for the language to be adopted by the HPC community.

After describing the approach we have adopted in adding pragma driven shared memory parallelism to Zig via supporting OpenMP loop directives and associated clauses in the compiler, we then undertook a performance analysis using NASA's NPB benchmark suite. We demonstrated that our approach delivers similar thread scaling to C and Fortran compilers, however it was observed that the runtime of the Zig benchmarks was lower than those of their Fortran counterparts. Given Fortran's lineage in scientific computing, these results surprised us and demonstrate that Zig is achieving its aim of delivering performance.

We believe the next step in driving Zig adoption in HPC will be to add support for profiling to the Zig compiler. At present, the Zig compiler is profiled using the Tracy library \cite{tracy}, however, the Zig interface to this library is part of the compiler itself and is not available to be used in applications. Modifying the compiler to automatically instrument applications with the calls to this library, providing functionality similar to that of gprof, would be worthwhile. Furthermore, enhancing interoperability between Zig and Fortran is important and would enable the integration of Zig into existing, large, Fortran codebases which are commonplace in HPC. Whilst in this paper we have demonstrated that such an integration is feasible, this requires additional engineering effort inside the compiler, and potential extension of the Fortran standard, to ensure that such an approach is reliable and consistent.

In summary, we conclude that the combination of performance and safety that is delivered by the Zig programming language has potential for it to be used for HPC workloads. By enhancing the compiler to support OpenMP loop directives, we have provided the capability for pragma driven shared memory parallelism in Zig and have demonstrated that scaling is comparable to other languages. Moreover, for HPC workloads performance matches, or sometimes exceeds, these other languages too.

\section*{Acknowledgment}
The authors acknowledge EPCC at the University of Edinburgh and EPSRC who have funded this work (EP/T517884/1) and the ExCALIBUR xDSL project (EP/W007940/1). This work used the ARCHER2 UK National Supercomputing Service (https://www.archer2.ac.uk). For the purpose of open access, the author has applied a Creative Commons Attribution (CC BY) licence to any Author Accepted Manuscript version arising from this submission.

\bibliographystyle{IEEEtran}
\bibliography{references.bib}

\end{document}